\documentclass[twocolumn,showpacs,preprintnumbers,amsmath,amssymb]{revtex4}

\usepackage{graphicx}
\usepackage{dcolumn}
\usepackage{bm}

\begin{document}

\preprint{S. J. MacLeod et al.}

\title{The role of background impurities in the single particle relaxation lifetime of a two-dimensional electron gas}

\author{S. J. MacLeod}
    \thanks{These authors contributed equally to this work}
\author{K. Chan}
    \thanks{These authors contributed equally to this work}
\author{T. P. Martin}
\author{A. R. Hamilton}
    \email{alex.hamilton@unsw.edu.au}
\author{A. See}
\author{A. P. Micolich}
    \affiliation{School of Physics, The University of New South Wales, Sydney, NSW 2052, Australia}
\author{M. Aagesen, P. E. Lindelof}
 \affiliation{Nano-science center, University of Copenhagen, Universitetsparken 5, DK-2100 Copenhagen, Denmark}
\date{\today}

\begin{abstract}
We re-examine the quantum $\tau_q$ and transport $\tau_t$ scattering lifetimes due to background impurities in two-dimensional systems. We show that the well-known logarithmic divergence in the quantum lifetime is due to the non-physical assumption of an infinitely thick heterostructure, and demonstrate that the existing non-divergent multiple scattering theory can lead to unphysical quantum scattering lifetimes in high quality heterostructures. We derive a non-divergent scattering lifetime for finite thickness structures, which can be used both with lowest order perturbation theory and the multiple scattering theory. We calculate the quantum and transport lifetimes for electrons in generic GaAs-AlGaAs heterostructures, and find that the correct `rule of thumb' to distinguish the dominant scattering mechanisms in GaAs heterostructures should be $\tau_{t}/\tau_{q} \lesssim 10$ for background impurities and $\tau_{t}/\tau_{q} \gtrsim 10$ for remote impurities.
Finally we present the first comparison of theoretical results for $\tau_q$ and $\tau_t$ with experimental data from a GaAs 2DEG in which only background impurity scattering is present. We obtain excellent agreement between the calculations and experimental data, and are able to extract the background impurity density in both the GaAs and AlGaAs regions.
\end{abstract}

\maketitle

\section{Introduction}\label{sec:intro}

Over the past four decades, an enormous effort has been dedicated to developing two-dimensional electron gases (2DEGs) engineered at the interface of semiconductor heterostructures, and optimising their low temperature transport properties~\cite{ando,PfeifferReview}. It is now possible to achieve low temperature mobilities in excess of $10^7 \textrm{cm}^2/\textrm{Vs}$ in high quality GaAs-AlGaAs heterostructures~\cite{Pfeiffer1989,Umansky1997}, and there is continued interest in further improving 2DEG mobilities~\cite{Umansky2009}. Phenomena that are only observed in ultra-high mobility 2DEGs include anisotropies, stripe and bubble phases in higher Landau levels~\cite{LillyPRL99}, microwave induced resistance oscillations~\cite{Mani2004}, and even denominator fractional quantum Hall states~\cite{WillettPRL87}. The latter have attracted considerable interest due to proposals for topological quantum computers that use the possible non-abelian nature of the $\nu=5/2$ fractional quantum Hall state~\cite{Sarma2005}. Since these even denominator fractional quantum Hall states are only observed in ultra-high mobility 2DEGs at very low temperatures ($T\lesssim100\textrm{mK}$), there is renewed interest in understanding the factors limiting the electron mobility in GaAs heterostructures~\cite{Hwang2008}.

The maximum attainable electron mobility is limited by phonon scattering, which cannot be avoided. However for low temperatures, $T<0.1$K, phonon scattering is negligible, and the mobility is limited by other mechanisms. In GaAs heterostructures these include interface roughness scattering, remote ionised impurity scattering, and background impurity scattering. Interface roughness scattering is not important in high quality heterointerfaces at low densities, so the low $T$ mobility is limited by ionised impurity scattering~\cite{GottingerEPL88,bock}. In most two-dimensional GaAs systems the carriers are introduced by modulation doping of the AlGaAs, and this modulation doping provides a significant source of remote ionised impurity scattering~\cite{Dingle78}. Remote ionised impurity scattering can be reduced with the use of large undoped AlGaAs `spacer' layers between the 2DEG and the modulation doping~\cite{StormerAPL81}, or eliminated entirely with accumulation mode devices in which the carriers are introduced electrostatically rather than through doping~\cite{Kane1993}.
Finally there are always `background impurities' incorporated in the crystal during the epitaxial growth process, and these will limit the mobility in both the very cleanest modulation doped samples~\cite{Umansky1997} and in electrostatically doped samples.

In practice, experimentally determining the factors limiting the mobility in a particular sample is non-trivial, since many scattering mechanisms may be acting together. There are two key experimental parameters available to try and separate the different scattering mechanisms: The transport scattering time $\tau_{t}$, obtained from the conductivity, and the single-particle relaxation time (also known as the quantum lifetime) $\tau_{q}$, obtained from the Shubnikov-de Haas oscillations. For short range isotropic scattering the two times are equivalent (e.g. in silicon MOSFETs), but for long range Coulomb interactions (e.g. from modulation doping) $\tau_t$ and $\tau_q$ are quite different~\cite{Harrang,Das,Coleridge}. The difference lies in the fact that $\tau_q$ counts all scattering events, while $\tau_{t}$ is weighted towards large-angle scattering events that cause a significant change in the momentum. By comparing the measured $\tau_t$ and $\tau_q$ with numerical calculations it is possible to determine the nature of the predominant scattering mechanism in a 2DEG at low temperatures~\cite{Das, Kearney, Harrang, Coleridge}. This is particularly important for background impurity scattering in ultra-high mobility 2D systems, since the impurity levels are so low that they cannot be measured by direct tools such as deep level transient spectroscopy.

While the transport scattering rates can be calculated using first order perturbation theory for a variety of scattering mechanisms, direct comparison between experimental measurements and theoretical calculations of $\tau_t$ and $\tau_q$ in high mobility samples have been limited by a problem in calculating the single-particle relaxation time for background impurity scattering: There is no lowest order result for the single-particle relaxation time for homogeneous background doping due a divergent integral~\cite{GoldPRB88}. In a series of papers, Gold and G\"{o}tze extended the theoretical formalism to include multiple scattering effects in both the transport and single-particle cases~\cite{GoldPRB88, goldgotze, goldgotze2}. This made it possible to calculate the `renormalized' single-particle lifetime for scattering from homogeneous background impurities~\cite{GoldPRB88}.

In this paper we re-examine the scattering lifetime due to background impurities, and explicitly highlight how the logarithmic divergence in $\tau_q$ is due to the assumption  of an infinitely thick heterostructure in Refs.~\cite{GoldPRB88,davies}. We derive a non-divergent scattering lifetime for a finite thickness heterostructure, both in the lowest order perturbation theory approach and in the multiple scattering theory. Comparing these two approaches shows how the existing multiple scattering theory can lead to inaccurate scattering lifetimes in high quality heterostructures. Finally we present the first comparison of theoretical results for $\tau_q$ and $\tau_t$ with experimental data from a GaAs 2DEG with only background impurity scattering. We obtain excellent agreement between the calculations and experimental data, and are able to extract the background impurity density in both the GaAs and AlGaAs regions.

The remainder of this paper is structured as follows: In Section II we review the standard concepts and expressions for ionised impurity scattering in a 2DEG, with special emphasis on homogenous background scattering, and introduce the correct expression for background impurity scattering in a finite thickness sample. In Section III we evaluate these expressions for background impurity scattering, and compare the different approaches to calculating the single particle and quantum lifetimes. In Section IV we compare our results with experimental data, followed by conclusions in Section V.

\section{Theory}
\begin{figure}[h]
\includegraphics[width=8cm]{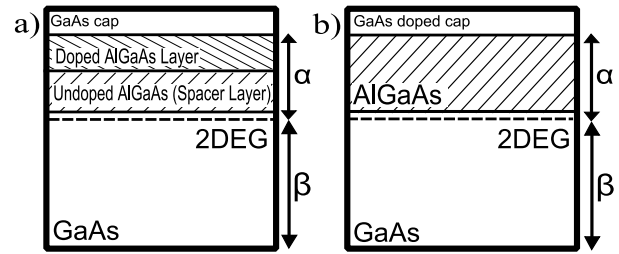}
\caption{Schematic of GaAs-AlGaAs heterostructures. a) Modulation doped heterostructure, in which carriers are introduced at the GaAs-AlGaAs interface by remote dopants, and b) `induced' field effect structure in which carriers are introduced by applying a bias to the heavily doped GaAs cap that acts as a metal gate.}
\label{fig:1 (induced device)}
\end{figure}

We restrict our analysis to 2DEGs at $T=0$, and concentrate on re-examining how the scattering lifetimes are calculated for background impurity scattering, since other scattering mechanisms are readily treated with first order perturbation theory~\cite{GoldPRB88,davies}.
Most GaAs-AlGaAs heterostructures share a common layer structure, as shown in Fig.~\ref{fig:1 (induced device)}: a GaAs substrate/buffer with AlGaAs layer(s) above, and the 2DEG accumulated at the heterointerface. The AlGaAs layer may contain remote ionised impurities, as in the conventional modulation doped HEMT shown in Fig.~\ref{fig:1 (induced device)}(a), or may be undoped as in `induced' field effect devices (Fig.~\ref{fig:1 (induced device)}(b)). The heterostructure in Fig.~\ref{fig:1 (induced device)}(b) has no modulation doping to form the 2DEG -- instead the degenerately doped GaAs cap acts as a metallic gate, and carriers are induced by electrostatic doping with a gate bias~\cite{Solomon1984, Kane1993}. For the purposes of evaluating background impurity scattering, both types of heterostructure can be treated as consisting essentially of two layers: A GaAs substrate below the 2DEG of thickness $\beta$, and AlGaAs layers on top of thickness $\alpha$~\cite{footnote1}.

\subsection{Background Impurity Scattering in the Lowest Order Theory}
In the lowest order of approximation, the transport and single-particle scattering lifetimes $\tau_{t}$ and $\tau_{q}$ are calculated at $T=0$ by integrating the scattering potential $|U(q)|^{2}$ with respect to the scattered wavevector $q$:
\begin{eqnarray}
\frac{1}{\tau_{t}}&=&\frac{m^{*}}{\pi\hbar^{3}k^{2}_{F}}\int^{2k_{F}}_{0}\frac{|U(q)|^{2}}{\epsilon(q)^{2}}\frac{q^2}{\sqrt{4k^{2}_{F}-q^{2}}}\, dq \label{eqn:lowerordertaut}\\
\frac{1}{\tau_{q}}&=&\frac{2m^{*}}{\pi\hbar^{3}}\int^{2k_{F}}_{0}\frac{|U(q)|^{2}}{\epsilon(q)^{2}}\frac{1}{\sqrt{4k^{2}_{F}-q^{2}}}\, dq \label{eqn:lowerordertauq}
\end{eqnarray}
Equations (\ref{eqn:lowerordertaut}) and (\ref{eqn:lowerordertauq}) can be derived via Fermi's Golden Rule or by invoking the second Klauder approximation to the self-energy $\Sigma(\mathbf{k}, E)$, where the formalism is presented in~\cite{GoldPRB88}. Here $m^{*}$ is the electron effective mass (taken as $m^{*}=0.067m_{e}$ for GaAs) and $k_{F}$ is the Fermi wave vector, which in two-dimensional reciprocal space is a measure of the radius of the mass-shell at $T=0$. We use the Thomas-Fermi approximation to the dielectric function $\epsilon(q)$:
\begin{eqnarray}
\epsilon(q)=1+\frac{1}{q}\frac{m^{*}}{\pi\hbar^{2}}\frac{e^2}{2\epsilon\epsilon_{0}}\frac{b(8b^{2}+9bq+3q^2)}{8(b+q)^{3}}
\end{eqnarray}
The dielectric function includes the form factor $b(8b^{2}+9bq+3q^2)/8(b+q)^{3}$ to account for the finite width of the 2DEG, where $b$ is a variational parameter and $1/b$ defines the thickness of the 2DEG~\cite{Fang}.

Equations (\ref{eqn:lowerordertaut}) and (\ref{eqn:lowerordertauq}) show that the magnitudes of $\tau_{t}$ and $\tau_{q}$ are related to the scattering potential $|U(q)|^2$, which is characterized by both the type of disorder and geometry of the quantum well that confines the 2DEG. To calculate the scattering from homogenous background impurities it is necessary to divide the bulk sample into infinitesimal layers, and treat each layer as a $\delta$-doped layer of remote ionized impurities at a distance $s$ from the 2DEG. The scattering potential due to one of these $\delta$-doped layers is:
\begin{eqnarray} \label{eqn:RII_scattering_Uq}
|U(q)|^{2}_{\text{RII}}=N_{i}\left(\frac{e^2}{2\epsilon\epsilon_{0}q}\right)^{2}e^{-2q|s|}F(q)
\end{eqnarray}
Here $N_{i}$ is the two-dimensional impurity density in the $\delta$-doped layer and $F(q)$ is the form factor that accounts for the finite thickness of the 2DEG.  There are two form factors, $F_{\text{AlGaAs}}(q)$ and $F_{\text{GaAs}}(q)$, for scattering sites located in the AlGaAs and GaAs regions respectively~\cite{footnote2}:
\begin{eqnarray} \label{eqn:AlGaAsFF}
F_{\text{AlGaAs}}(q)=\frac{1}{(1+q/b)^{6}} 
\end{eqnarray}
\begin{eqnarray}\label{eqn:GaAsFF}
F_{GaAs}(q)=\frac{2+24(q/b)^{2}+48(q/b)^{3}+18(q/b)^{4}+3(q/b)^{5}}{2(1+q/b)^{6}}
\nonumber
\end{eqnarray}
The difference between the two form factors is due to the overlap between the wave function of the 2DEG and the background impurities in the GaAs region.

The total scattering potential is obtained by summing the contribution from all of the delta-layers:
\begin{eqnarray} \label{eqn:BackgroundInfiniteSum}
|U(q)|^{2}_{\text{BG}}&=&\int^{\infty}_{0}|U(q)|^{2}_{\text{RII AlGaAs}}\,dz+ \nonumber \\
&& \int^{0}_{-\infty}|U(q)|^{2}_{\text{RII GaAs}}\,dz
\end{eqnarray}
 to obtain:
\begin{eqnarray} \label{eqn:BackgroundSP}
|U(q)|^{2}_{\text{BG}}=\frac{N_{B}}{2q}\left(\frac{e^2}{2\epsilon\epsilon_{0}q}\right)^{2}[F_{\text{AlGaAs}}(q)+F_{\text{GaAs}}(q)]
\end{eqnarray}
The three-dimensional background impurity density is now defined by $N_{B}$.

Equations~(\ref{eqn:lowerordertaut})--(\ref{eqn:BackgroundSP}) represent the standard lowest order expression for scattering from background impurities. Eqn.~(\ref{eqn:lowerordertaut}) is well behaved, but Eqn.~(\ref{eqn:lowerordertauq}) diverges as $q \rightarrow 0$ and cannot be evaluated.

The divergence in $\tau_q$ for background impurity scattering arises from the infinite limits in Eqn.~(\ref{eqn:BackgroundInfiniteSum}), which physically corresponds to an infinite sample, containing an infinite number of charged impurities located far from the 2DEG. These charged background impurities, most of which are located far from the 2DEG, lead to a divergence in the small angle (small $q$) scattering rate due to the $e^{-2q|s|}$ term in Eqn.~(\ref{eqn:RII_scattering_Uq}), and hence to a divergence in $\tau_{q}$. However the transport lifetime is convergent since it is less sensitive to small angle scattering events. The non-physical assumption of an infinitely thick sample has been a common assumption in previous calculations of background impurity scattering~\cite{gold2, gold3,gold4, Kearney,davies}.

To solve the problem of the logarithmic divergence in the single-particle scattering rate, it is tempting to simply modify the lower limit of the integral in Eqn.~(\ref{eqn:lowerordertauq}) and replace it with the uncertainty of the 2D scattered wave vector $q$. However, there are several problems with this approach. First of all, the integrand in Eqn.~(\ref{eqn:lowerordertauq}) contributes most strongly near the two limits of integration. Thus the result of the integral is extremely sensitive to any modifications of these two limits.  Secondly one requires a precise knowledge of all the physical constraints that specify the uncertainty in $q$ regarding the system of interest.  The uncertainty in $q$ will depend on some of the same parameters that affect the scattering rate that one intends to calculate. Only an approach which addresses the problem self-consistently would allow one to be confident of the results.  Thirdly even if one can remove the logarithmic divergence with this approach, there still exists the assumption that the heterostructure is infinitely thick.  This assumption is not only physically unsound, but also produces inaccurate results as we demonstrate in Section III.

\subsection{The Single-Particle Lifetime in the Higher Order Theory}
The conventional way around the logarithmic divergence in the integral for $\tau_q$ is to use the multiple scattering theory developed by Gold to calculate the renormalized single-particle scattering lifetime $\tau_{qr}$~\cite{GoldPRB88}.
The starting point for incorporating multiple scattering effects into the single-particle lifetime is to use the single-particle Green's function with the mass-shell and third Klauder approximations~\cite{GoldPRB88, Klauder}. Taking into account only electron-impurity interactions (\emph{i.e.} neglecting electron-electron interactions) one arrives at the following expression for $\tau_{qr}$:
\begin{eqnarray} \label{eqn:tauqrDoubleInt}
\frac{1}{\tau_{qr}}&=&\frac{2m^{*}}{\hbar^3\pi^2}\int^{\infty}_{0} \int^{\frac{\pi}{2}}_{-\frac{\pi}{2}}
q\frac{|U(q)|^2}{\epsilon(q)^{2}}  \nonumber \\
& & \times
\frac{m^{*}/\hbar\tau_{qr}}{(q^2+2qk_{F}\text{sin}\varphi)^2+(m^{*}/\hbar\tau_{qr})^2} \, d\varphi \, dq
\end{eqnarray}

Eqn.~(\ref{eqn:tauqrDoubleInt}) is a self-consistent equation with a double integral that must be solved recursively to obtain $\tau_{qr}$. This is a somewhat involved, and computationally intensive, operation and so has rarely (if ever) been used to model experimentally obtained single-particle scattering lifetimes. It is possible to simplify equation (\ref{eqn:tauqrDoubleInt}) by carrying out analytical approximations as in Ref.~\cite{GoldPRB88}, but this introduces significant deviations from the exact result, particularly for low disorder GaAs-AlGaAs heterostructures~\cite{macleod2007}.

Kearney \emph{et al.} showed that the double integral of equation~(\ref{eqn:tauqrDoubleInt}) can be reduced to a single integral through contour integration~\cite{Kearney}:
\begin{eqnarray} \label{eqn:tauqrSingleInt}
\frac{1}{\tau_{qr}}=\frac{m^*}{\hbar^3\pi}\int^{\infty}_{0} \frac{|U(q)|^2}{\epsilon(q)^{2}}
\left(\frac{\hbar\tau_{qr}}{m^*}\right)^{\frac{1}{2}}\frac{e^{\theta(q)/2}}{\text{cosh}\theta(q)} \, dq
\end{eqnarray}
where $\theta(q)$ is given by:
\begin{eqnarray} \label{eqn:theta_q}
\theta(q)=\text{sinh}^{-1}\left[\frac{(m^*/\tau_{qr})^2-q^4+4k^{2}_{F}q^2}{2(m^*/\tau_{qr})q^2}\right]
\nonumber \end{eqnarray}

\begin{figure}[h]
\includegraphics[width=8cm]{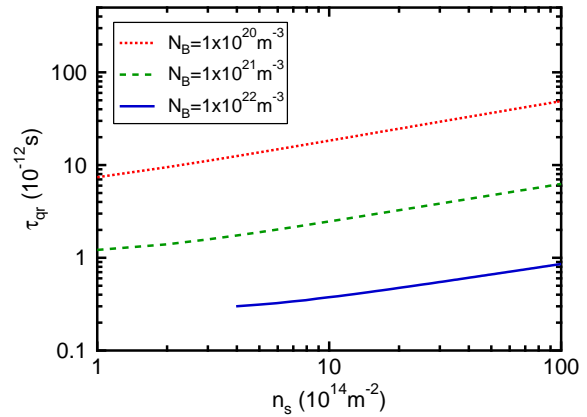}
\caption{$\tau_{qr}$ due to background impurity scattering for an infinitely thick sample, plotted against the electron density $n_{s}$ of the 2DEG. The doping densities in the GaAs and AlGaAs regions are taken to be the same,  $N_B=N_{B1}=N_{B2}$. Results are plotted only in the regime where the mass shell approximation is valid, $E_F>\hbar/\tau_{qr}$.}
\label{fig:tauqrVscarrierdensityINF}
\end{figure}

Equation~(\ref{eqn:tauqrSingleInt}) allows the renormalised single particle relaxation time for background scattering
for an infinitely thick heterostructure to be calculated efficiently (in a few minutes) using a standard personal computer. To demonstrate this we evaluate $\tau_{qr}$ as a function of the carrier density for a generic 2DEG in a GaAs-AlGaAs heterostructure (c.f. Fig.~\ref{fig:1 (induced device)}), for three different background impurity densities, as shown in Fig.~\ref{fig:tauqrVscarrierdensityINF}. The scattering lifetime decreases with decreasing carrier density, although it tends to saturate slightly at low densities. Unlike the lowest order scattering times $\tau_t$ and $\tau_q$, the self-consistent nature of $\tau_{qr}$ means that it is not a simple linear function of the background impurity density: Increasing $N_B$ by a factor of 10 does not simply reduce $\tau_{qr}$ by a factor of 10.

\subsection{A Physically Realistic Scattering Potential for Homogeneous Background Impurities}
Although Fig.~\ref{fig:tauqrVscarrierdensityINF} provides convergent results for the single-particle lifetime due to background impurity scattering, it is instructive to identify the source of the divergence that caused problems for the lower order theory in the first place. It is a simple exercise to correct the background scattering potential and set a realistic thickness for the heterostructure, with finite bounds on the integrals in Eqn.~(\ref{eqn:BackgroundInfiniteSum}). This leads to the potential:
\begin{eqnarray} \label{eqn:FiniteBackgroundSP}
|U(q)|^{2}_{\text{BG}}&=&\left(\frac{e^2}{2\epsilon\epsilon_{0}q}\right)^{2}\frac{1}{2q}\bigg[N_{B1}F_{\text{AlGaAs}}(q)(1-e^{-2q\alpha}) \nonumber\\
&&+N_{B2}F_{\text{GaAs}}(q)(1-e^{-2q\beta})\bigg]
\end{eqnarray}
Here $\alpha$ and $N_{B1}$ are the thickness and background doping levels in the AlGaAs layer above the 2DEG, and $\beta$ and $N_{B2}$ are the thickness and background doping levels in the GaAs layer below the 2DEG. Using the scattering potential in Eqn.~(\ref{eqn:FiniteBackgroundSP}) the lowest order single particle relaxation time $\tau_q$ can now be calculated without divergence. This scattering potential can also be used in Eqn.~(\ref{eqn:tauqrSingleInt}) to calculate the renormalised $\tau_{qr}$ for a sample with finite thickness.

\section{Evaluation of theoretical scattering times}
Having removed the divergence in the calculation of the single particle scattering lifetime, we can now evaluate and compare the various lifetimes for different sample thicknesses and doping levels.

\subsection{A Comparison of $\tau_t$, $\tau_{q}$ and $\tau_{qr}$}
\begin{figure}[h]
\includegraphics[width=8.6cm]{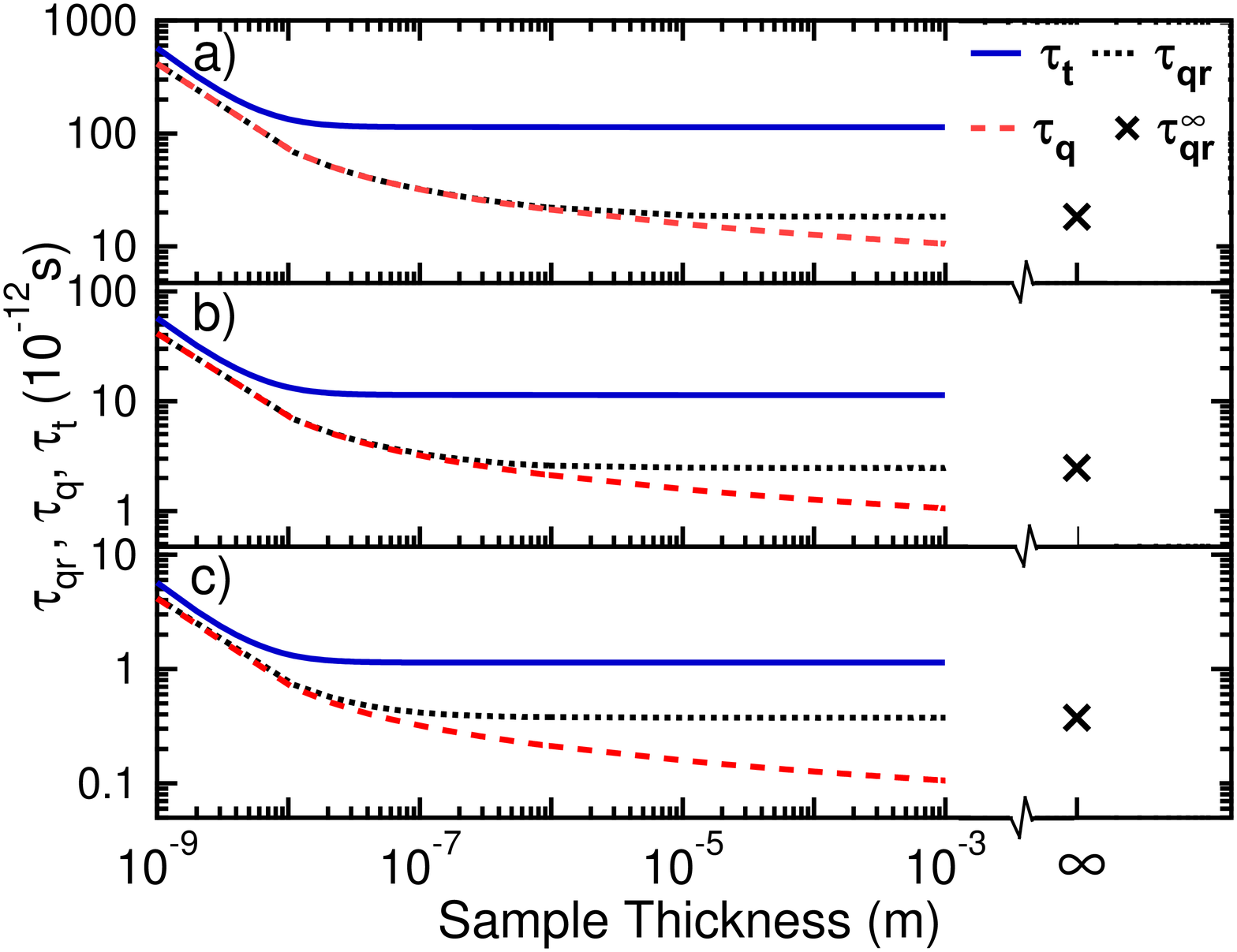}
\caption{$\tau_{t}$, $\tau_{q}$ and $\tau_{qr}$ were calculated with the background scattering potential for a finite-width sample and are plotted against the sample thickness. $N_{B}$ is varied between $10^{20}\textrm{m}^{-3}$ and $10^{22}\textrm{m}^{-3}$ in panels (a)-(c), and $n_{s}$ is fixed at $n_{s}=10^{15}m^{-2}$ (similar results are achieved for electron densities of $10^{14}\textrm{m}^{-2}$ and $10^{16}\textrm{m}^{-2}$). Crosses mark $\tau^{\infty}_{qr}$ calculated with the background scattering potential derived for an infinitely thick sample~\cite{gold2}. It can be seen that $\tau^{\infty}_{qr}$ underestimates the scattering lifetime for thin structures.}
\label{fig:tau_thickness}
\end{figure}

In Figure~\ref{fig:tau_thickness} we plot the scattering lifetimes evaluated for the generic GaAs-AlGaAs heterostructure of Fig.~\ref{fig:1 (induced device)} as a function of sample thickness, for three different background impurity concentrations. For these calculations we have taken the background impurity levels in the GaAs and AlGaAs regions to be the same $N_B=N_{B1}=N_{B2}$. We also set the thickness of the two regions to be the same $\alpha=\beta$, so that the total sample thickness is $s=\alpha+\beta$. The electron effective mass is taken as $0.067 m_e$, and the dielectric constant of both GaAs and AlGaAs as $12.7\varepsilon_0$.

The solid line in Fig.~\ref{fig:tau_thickness} shows the transport relaxation time $\tau_t$ as a function of sample thickness, for three different levels of the background doping density $N_B$. As the sample is made thicker there are more and more background impurities for electrons to scatter from, and $\tau_t$ decreases. Above $\sim100$~nm the scattering rate $\tau_t$ saturates, since background impurities located more than 100~nm from the 2DEG essentially act as remote impurities, predominantly causing small angle scattering that does not affect the momentum. 

The dashed line in Fig.~\ref{fig:tau_thickness} shows the lower order single-particle lifetime $\tau_{q}$, calculated using Eqn.~(\ref{eqn:FiniteBackgroundSP}). The single particle lifetime $\tau_q$ now produces convergent results, and like $\tau_t$ it decreases as the sample becomes thicker (larger $s$). The single particle scattering time is always shorter than the equivalent transport scattering time, since $\tau_q$ counts all scattering events and $\tau_t$ is weighted towards large angle scattering events. However $\tau_q$ exhibits the non-physical behaviour of continually decreasing as the thickness is increased, so that even impurities located at an almost infinite distance contribute to the scattering, and the single particle scattering rate diverges as $s \rightarrow \infty$. Even though $\tau_q$ can now be calculated for real samples, it cannot safely be used to obtain the ratio $\tau_t/\tau_q$ for comparison with experiments, since $\tau_q$ is always sensitive to the thickness of the sample, and it overestimates the scattering compared to the higher order $\tau_{qr}$.
Note that since $\tau_t$ and $\tau_q$ are calculated to lowest order, the scattering rate is directly proportional to $N_B$, so the solid and dashed curves in the three panels of Fig.~\ref{fig:tau_thickness} differ only by a numerical prefactor ($N_B$).

The crosses in Fig.~\ref{fig:tau_thickness} show the renormalised single particle scattering time $\tau_{qr}^\infty$, calculated using the higher order multiple scattering theory in Eqn.~(\ref{eqn:tauqrSingleInt}). As originally defined by Gold~\cite{GoldPRB88}, this scattering time is evaluated for an infinitely thick sample, yet produces a convergent result. Since the renormalised scattering time is calculated iteratively ($\tau_{qr}$ exists on both the left and right sides of Eqn.~(\ref{eqn:tauqrSingleInt})), $\tau_{qr}^\infty$ is not a simple function of $N_B$; increasing $N_B$ tenfold does not necessarily increase $\tau_{qr}$ by a factor of ten.

It is instructive to examine how the renormalised scattering time behaves for finite thickness samples, using the finite thickness scattering potential (Eqns.~(\ref{eqn:tauqrSingleInt}) and (\ref{eqn:FiniteBackgroundSP})). The dotted lines in Fig.~\ref{fig:tau_thickness} show $\tau_{qr}$ calculated as a function of sample thickness. The renormalised single-particle lifetime saturates to a finite value, $\tau_{qr}^\infty$, as the thickness of the sample is increased, although it takes longer to saturate than the transport scattering time. The dotted lines in Figs.~\ref{fig:tau_thickness}(a-c) also show that as the background impurity density $N_B$ is increased, $\tau_{qr}$ saturates at a smaller sample thickness. Physically this indicates that the `dirtier' a sample is, the less effect impurities situated away from the 2DEG have on the scattering of electrons in the 2DEG.

The calculations of $\tau_{qr}$ for finite thickness samples also highlight a problem with clean samples that are thinner than $\sim 1\mu$m, such as heterostructure epilayers:  $\tau_{qr}$ hasn't yet saturated to $\tau_{qr}^\infty$, so the standard assumption of an infinite heterostructure thickness~\cite{gold2, gold3,gold4, Kearney} results in an overestimation of the renormalised single-particle scattering rate $1/\tau_{qr}^\infty$. This result clearly demonstrates that the scattering potential $U(q)$ must take into account the finite thickness of the heterostructure for accurate comparison with experimental data. \\

\begin{figure}[h]
\includegraphics[width=8cm]{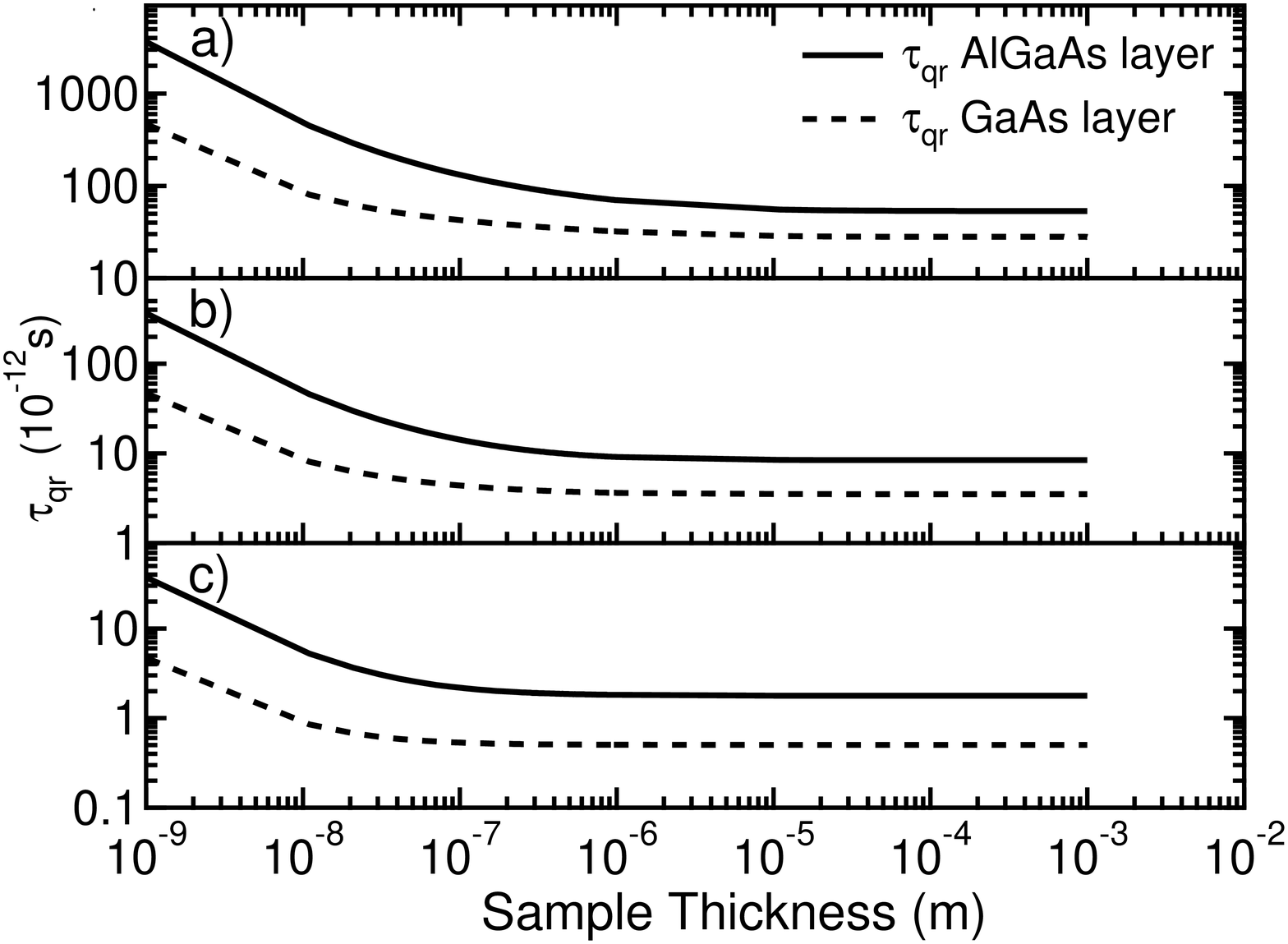}
\caption{$\tau_{qr}$ due to background impurity scattering plotted separately for the AlGaAs and GaAs regions as a function of thickness, showing that most of the scattering originates from the GaAs layer. In panels (a)-(c) $n_{S}$ is fixed at $n_{s}=10^{15}m^{-2}$ and N$_{B}$ is varied between $10^{20}\textrm{m}^{-3}$ and $10^{22}\textrm{m}^{-3}$.}
\label{fig:1b (tqr thickness-sep regions)}
\end{figure}

Since the GaAs and AlGaAs layers in a real sample are generally not of equal thickness (typically the AlGaAs region is $\sim100$nm and the GaAs region is $\sim200\mu\text{m}$), it is useful to investigate how quickly $\tau_{qr}$ saturates in each layer as a function of the layer thickness. However since $\tau_{qr}$ is calculated recursively and self-consistently through Eqn.~(\ref{eqn:tauqrSingleInt}), Matthieson's rule no longer holds, and the scattering rates for the two regions cannot be calculated independently of each other. We therefore have to calculate $\tau_{qr}$ for the complete sample, and then determine the fraction of the total scattering caused by each layer. To do this $\tau_{qr}$ was calculated for GaAs and AlGaAs layers having the same thickness and doping levels, $\alpha=\beta$. After $\tau_{qr}$ was obtained for the entire sample, this value of $\tau_{qr}$ was inserted into the right hand side of Eqn.~(\ref{eqn:tauqrSingleInt}), with the appropriate choice of $U(q)$ for either the AlGaAs or GaAs region, to obtain the scattering from each layer individually. The results are plotted in Fig.~\ref{fig:1b (tqr thickness-sep regions)} as a function of the layer thickness $s=\alpha+\beta$. The calculated scattering times show that there are two significant differences between the single-particle lifetime for the AlGaAs and GaAs regions: (i) Most of the contribution to the total scattering lifetime $\tau_{qr}$ comes from impurities in the GaAs region; (ii) The scattering contribution from the GaAs saturates faster than that from the AlGaAs region as $s$ is increased. Physically this means that most of the scattering in the GaAs region comes from impurities close to the channel ($s\lesssim10\text{nm}$), whereas in the AlGaAs region impurities located a considerable distance from the 2DEG can still cause appreciable scattering. The results in Fig.~~\ref{fig:1b (tqr thickness-sep regions)} again highlight the problem with the conventional assumption of an infinite sample thickness when calculating $\tau_{qr}$, since in most  heterostructures the AlGaAs epilayer is only a few hundred nm thick yet $\tau_{qr}^{AlGaAs}$ doesn't saturate to the infinite limit until $s \gtrsim 1\mu\text{m}$.

\subsection{Comparison of $\tau_{t}/\tau_{qr}$ and $\tau_{t}/\tau_{q}$}

\begin{figure}[h]
\includegraphics[width=8cm]{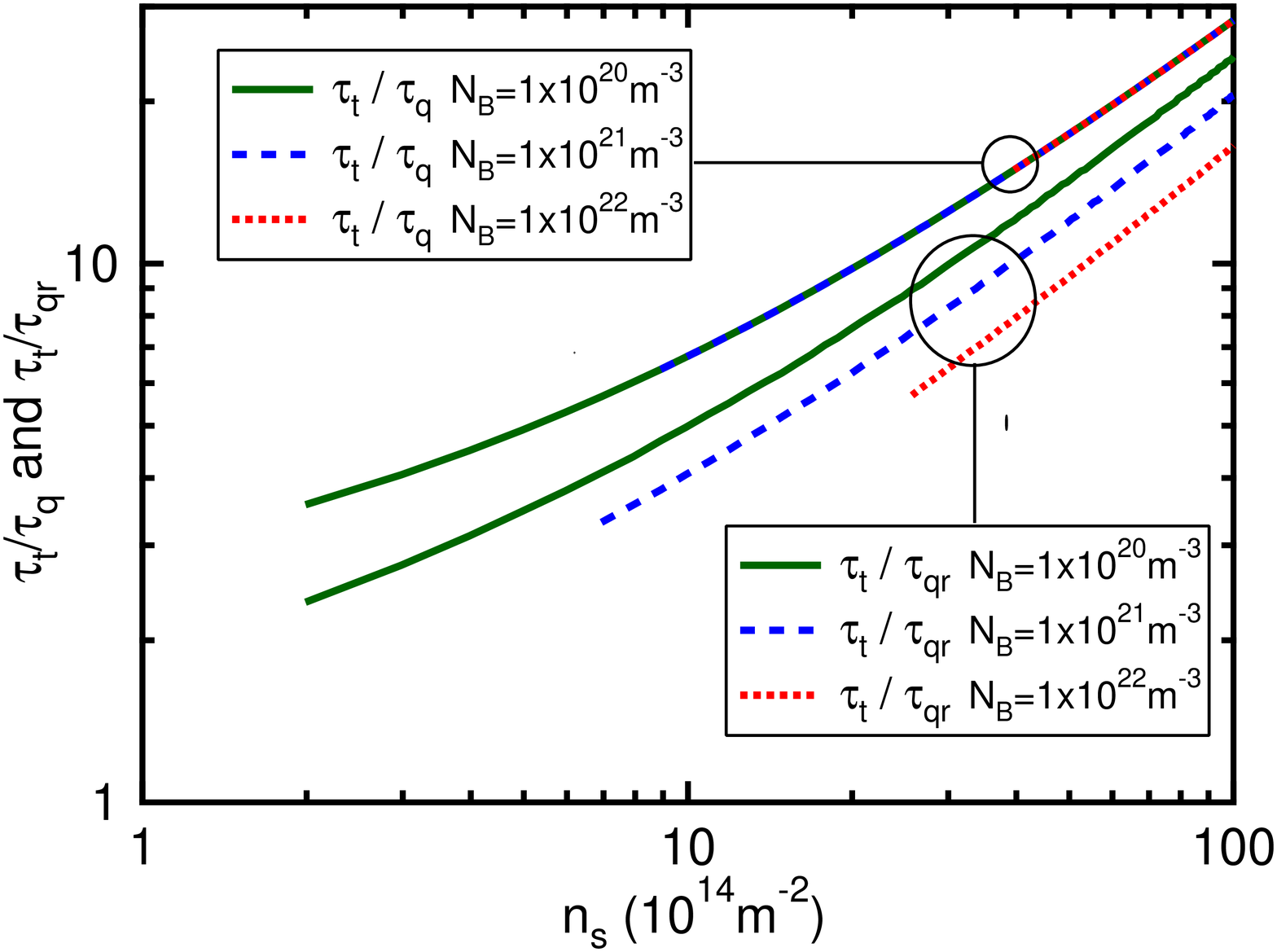}
\caption{The ratio $\tau_{t}/\tau_{q}$ and $\tau_{t}/\tau_{qr}$ as a function of the 2DEG carrier density $n_s$, showing that the ratio can be significantly larger than 1 for background impurity scattering. Calculations were performed with the background scattering potential for a finite-width sample ($\alpha=100\text{nm}$ and $\beta=300\mu\text{m}$), for three different values of the background impurity density ($N_B=N_{B1}=N_{B2}$).}
\label{fig:taut-on-tauq}
\end{figure}

It is widely assumed that the ratio of the transport to the single particle lifetime $\tau_t / \tau_q$ can be taken as an indication of the dominant scattering mechanism~\cite{Das, Kearney, Harrang, Coleridge}. The conventional wisdom is that $\tau_t / \tau_q$ should be close to 1 for background impurity scattering and much larger than 1 for remote ionised impurity scattering. To test if this is correct we plot in Fig.~\ref{fig:taut-on-tauq} both $\tau_{t}/\tau_q$ and $\tau_t/\tau_{qr}$ calculated for background impurity scattering only, as a function of the 2D carrier density. We take the `generic' thicknesses of the heterostructure to be $\alpha=100\text{nm}$ for the AlGaAs region and $\beta=300\mu\text{m}$ for the GaAs region,  with three different background doping levels ($N_{B1}=N_{B2}$) varied from $10^{20}-10^{22}\text{m}^{-3}$.

In Fig.~\ref{fig:taut-on-tauq} the calculation is performed for a finite sample, so the lowest order single particle scattering time $\tau_q$ is convergent and both $\tau_{t}/\tau_{q}$ and $\tau_{t}/\tau_{qr}$ can be evaluated. There is no self-consistency in the lowest order theory (the impurity concentration is effectively just a prefactor in the integrals in Eqns.~(\ref{eqn:lowerordertaut}) and~(\ref{eqn:lowerordertauq})), so $\tau_{t}/\tau_{q}$ is insensitive to $N_B$. In contrast $\tau_{t}/\tau_{qr}$ decreases as $N_{B}$ is increased, indicating that there is proportionally more large angle scattering as the background impurity density is increased.

The key result of Fig.~\ref{fig:taut-on-tauq} is that even for background impurity scattering alone $\tau_{t}/\tau_{qr}$ can be as high as ~10, which is much larger than conventional wisdom would suggest. We suggest instead that the correct `rule of thumb' to distinguish the dominant scattering mechanisms in GaAs heterostructures should be $\tau_{t}/\tau_{qr} \lesssim 10$ for background impurity and $\tau_{t}/\tau_{qr} \gtrsim 10$ for remote ionised impurity scattering~\cite{macleod2007}.

\section{Comparison with experimental data}
We now compare our results with experimental data from a 2DEG in a GaAs-AlGaAs heterojunction. This sample is specially designed to only have background impurity scattering, to allow direct comparison between experiment and theory. The sample has no modulation doping to form the 2DEG -- instead carriers are induced by electrostatic doping with a gate bias (this type of sample is referred to as a Semiconductor Insulator Semiconductor FET, SISFET~\cite{Solomon1984}, or Heterostructure Insulated Gate FET, HIGFET~\cite{ZhuPRL03}). The sample consists of a 450$\mu\text{m}$ substrate, 1$\mu\text{m}$ GaAs, 160nm AlGaAs and 60nm GaAs cap. Devices were fabricated in a Hall bar geometry, and standard low-frequency a.c. magnetotransport measurements were performed at 1.4~K with an excitation of 100 $\mu\text{V}$.

\begin{figure}[h]
\includegraphics[width=8.6cm]{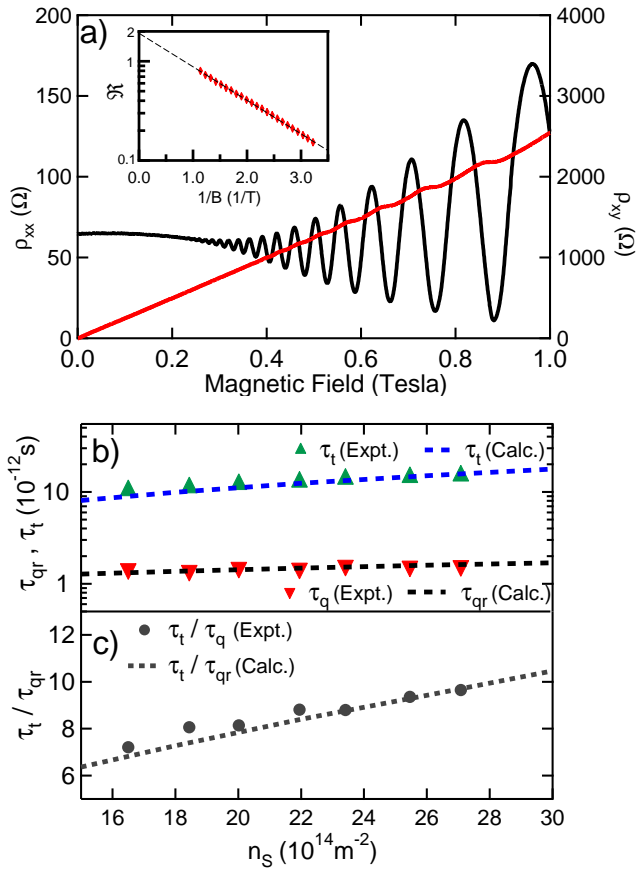}
\caption{(a) Longitudinal and Hall magneto-resistance of an `induced' GaAs 2DEG in a SISFET structure at $T=1.4$~K. Inset shows the Dingle plot analysis of the reduced resistivity $\Re$ as a function of the inverse magnetic field. (b,~c) Experimentally measured values of $\tau_{t}$, $\tau_{q}$ and the ratio $\tau_t / \tau_q$ compared with calculated scattering times. Experimental values are indicated by the solid symbols and the theoretically predicted trends are indicated by the dashed lines. For the theoretical curves $N_{B1}=6.25\times 10^{21}\textrm{m}^{-3}$ and $N_{B2}=1.92\times10^{21}\textrm{m}^{-3}$.}
\label{fig:TheoryExperimentComparison}
\end{figure}

Figure~\ref{fig:TheoryExperimentComparison}(a) shows magnetotransport data from sample NBI30/AS08N with a top-gate bias of 1.05~V, corresponding to a 2D carrier density of $2.02\times10^{11}\textrm{cm}^{-2}$. Carrier densities extracted from the low field Hall effect and the periodicity of the Shubnikov-de Haas (SdH) oscillations agreed to within 2\%. We extract the transport lifetime from the resistivity at $B=0$, and the quantum lifetime from a Lifshitz-Kosevitch analysis of the SdH oscillations~\cite{ColeridgePRB1991}. The inset to Fig.~\ref{fig:TheoryExperimentComparison}(a) shows a Dingle plot of the reduced amplitude of the SdH oscillations, defined as $\Re=\Delta\rho_{xx} / 2D(X)\rho_0$ where $D(X)=X/\sinh(X)$ and $X=2\pi^2k_B T/\hbar\omega_c$, as a function of inverse magnetic field. The data falls onto a straight line, with a intercept close to 2 at $1/B=0$, which is a characteristic of a `good' Dingle plot from which a reliable quantum scattering time can be extracted~\cite{ColeridgePRB1991}.

The extracted transport and single particle lifetimes are plotted as solid symbols in Figure~\ref{fig:TheoryExperimentComparison}(b) as a function of the 2D carrier density $n_s$. As expected, the scattering lifetimes increase with increasing $n_s$, with the transport lifetime showing a stronger density dependence than the quantum lifetime.
The dashed and solid lines show the scattering times calculated for background impurity scattering only. For these calculations the sample is modelled as consisting of two layers, a 160nm thick AlGaAs layer above the 2DEG and a 450$\mu$m GaAs buffer below it (although $\tau_{qr}$ has already saturated with a GaAs thickness of 1$\mu$m). The renormalised quantum lifetime $\tau_{qr}$ was calculated using the finite thickness scattering potential defined in Eqn.~(\ref{eqn:FiniteBackgroundSP}). The only fitting parameters were the background impurity density of the AlGaAs and GaAs regions, $N_{B1}$ and $N_{B2}$. It was only possible to achieve a good fit of the calculated scattering times to the experimental data when $N_{B1}$ and $N_{B2}$ were different, with $N_{B1}=6.25\times10^{21}\textrm{m}^{-3}$ and $N_{B2}=1.92\times10^{21}\textrm{m}^{-3}$ giving the best fit.
The higher background doping level $N_{B2}$ in the AlGaAs layer is consistent with previous studies showing AlGaAs has a higher background doping level than GaAs~\cite{clarke, Laihktman1990}. We note that the quality of the fits and fitting parameters for $\tau_{qr}$ and $\tau_t/\tau_{qr}$ were rather insensitive to the thickness of the GaAs region $\beta$, changing by less than 2\% in the range $1<\beta<450\mu$m, whereas $\tau_q$ was very sensitive to $\beta$ (not shown). This reinforces the need to use the renormalised quantum lifetime.

Fig.~\ref{fig:TheoryExperimentComparison}(b) shows the ratio of the transport to quantum scattering lifetimes for the SISFET sample. This device has no modulation doping, so that scattering is by background impurities only, yet the ratio of scattering lifetimes is of order 10, much larger than the conventional wisdom that $\tau_t/\tau_q \sim 1$ for background impurity scattering. This reinforces the rule of thumb introduced earlier, and highlights the need to perform rigorous modelling of experimental scattering times and the ratio $\tau_t/\tau_q$ in order to determine the limiting scattering mechanism in high quality 2D samples.

\section{Conclusions}
In this paper we have re-analysed the problem of background impurity scattering for 2DEGs in semiconductor heterojunctions. We have shown that current approaches to calculating the quantum lifetime due to background impurities either fail completely, or produce inaccurate results in high quality heterostructures at low electron densities -- precisely the area of interest for modern devices. We derived a non-divergent scattering lifetime for finite thickness structures, and have shown that this can be used both with the lowest order perturbation theory and the multiple scattering theory, although only the latter produces physically sensible results. We have found excellent agreement between  theoretical calculations and experimental measurements of the scattering times due to background impurities for a GaAs 2DEG in which only background impurity scattering is present. Although our analysis was presented for AlGaAs-GaAs systems, this approach is applicable to generic semiconductor heterostructures.

\begin{acknowledgments}
We thank Fred Green for many useful discussions, and Oleh Klochan for assistance with measurements.
This work was funded by the Australian Research Council through the Discovery Projects Scheme (DP0772946); ARH acknowledges an ARC APF grant.
\end{acknowledgments}


\begin{thebibliography}{36}
\bibitem[{\citenamefont{Footnote 1}(2009)}]{footnote1}
\bibinfo{note}{A thin GaAs capping layer is often present in modulation doped structures, to prevent oxidation of the AlGaAs, but this can be treated as part of the AlGaAs layer for the purposes of background impurity scattering}.
\bibitem[{\citenamefont{Footnote 2}(2009)}]{footnote2}
\bibinfo{note}{The form factors given in Eqn.~(\ref{eqn:AlGaAsFF}) differ from those given by Gold in Ref.~\cite{gold2} by a factor of 1/2. This factor of 1/2 comes from the integration of Eqn.~(\ref{eqn:BackgroundInfiniteSum}), and can either be accounted for in the form factors of Eqn.~(\ref{eqn:AlGaAsFF}) (as Gold does) or in the $|U(q)|^2_{\text{BG}}$ of Eqn.~(\ref{eqn:BackgroundSP}) (as we do)}.
\end{thebibliography}
\end{document}